\documentclass[iop]{emulateapj}
\usepackage{apjfonts}
\usepackage[usenames]{color}
\usepackage{hyperref}

\hypersetup{colorlinks=true,linkcolor=black,citecolor=blue,filecolor=cyan,urlcolor=magenta}

\newcommand{\kms}{km\,s$^{-1}$}

\newcommand{\cM}{{\cal{M}}}
\newcommand{\mM}{m\!-\!M}
\newcommand{\OOc}{\Omega/\Omega_{\rm C}}

\newcommand{\mybibitem}[3]{\bibitem[{#1}({#2})]{#3}}
\newcommand{\mybibthree}[4]{\bibitem[{#2}({#3}){#1}]{#4}}
\slugcomment{Published in ApJ, {\bf 846}, 22 (2017)}

\shorttitle{eMSTOs in intermediate-age star clusters}
\shortauthors{Goudfrooij et al.}

\begin{document}

\title{Extended Main Sequence Turnoffs in Intermediate-Age Star Clusters: \\
Stellar Rotation diminishes, but does not eliminate, Age Spreads\altaffilmark{1}}
\definecolor{MyBlue}{rgb}{0.3,0.3,1.0}
\author{Paul Goudfrooij$^2$, L\'eo Girardi$^3$ and Matteo Correnti$^2$} 
\affil{$^2$ Space Telescope Science Institute, 3700 San Martin
  Drive, Baltimore, MD 21218, USA;  \href{mailto:goudfroo@stsci.edu}{\color{MyBlue}goudfroo@stsci.edu},  {\color{MyBlue}{correnti@stsci.edu}} \\
 $^3$ Osservatorio Astronomico di Padova -- INAF, Vicolo dell'Osservatorio 5,
  I-35122 Padova, Italy; {\color{MyBlue}leo.girardi@oapd.inaf.it}}

\altaffiltext{1}{Based on observations with the NASA/ESA {\it Hubble
    Space Telescope}, obtained at the Space Telescope Science
  Institute, which is operated by the Association of Universities for
  Research in Astronomy, Inc., under NASA contract NAS5-26555} 




\begin{abstract}
Extended main sequence turn-off (eMSTO) regions are a common feature in 
color-magnitude diagrams of young and intermediate-age star clusters in the
Magellanic Clouds. The nature of eMSTOs remains debated in the
literature. The currently most popular scenarios are extended star formation
activity and ranges of stellar rotation rates. 
Here we study details of differences in MSTO morphology expected from spreads in
age versus spreads in rotation rates, using Monte Carlo simulations
with the Geneva {\sc syclist} isochrone models that include the effects of
stellar rotation. 
We confirm a recent finding of Niederhofer et al.\ that a distribution of
stellar rotation velocities yields an MSTO extent that is proportional to the 
cluster age, as observed. However, we find that stellar rotation yields
MSTO crosscut \emph{widths} that are generally smaller than observed
ones at a given age.  
We compare the simulations with high-quality \emph{Hubble Space Telescope}
data of NGC~1987 and NGC~2249, which are the two only relatively
massive star clusters with an age of $\sim$\,1 Gyr for which such data
is available. We find that the distribution of stars across the eMSTOs
of these clusters cannot be explained solely by a distribution of
stellar rotation velocities, unless the orientations of rapidly
rotating stars are heavily biased towards an equator-on configuration.
 Under the assumption of random viewing angles, stellar rotation can
 account for $\sim$\,60\% and $\sim$\,40\% of the observed FWHM widths
 of the eMSTOs of NGC~1987 and NGC~2249, respectively.  
In contrast, a combination of distributions of 
stellar rotation velocities \emph{and} stellar ages fits the observed
eMSTO morphologies very well.
\end{abstract}




\section{Introduction} 
\label{s:intro}

Star clusters were long thought to be formed in a single burst of star
formation that produced thousands to millions of stars with the same age and
chemical composition, gravitationally bound together in a common potential
well. However, several studies over the last $\sim$\,15 years have shown that
this traditional view oversimplified the situation. Many globular clusters 
(GCs) in the Local Group are now known to harbor multiple stellar populations
exhibiting different abundances of light elements \citep[see][for a recent
review]{grat+12}. One of the best known signatures of such multiple stellar
populations is the Na-O anticorrelation among stars within clusters, which is
known to exist in the great majority of globular clusters studied to date in
sufficient detail \citep[e.g.,][]{carr+10}. The material responsible for the light
element abundance variations is generally thought to accumulate in the cluster
during its early evolution through slow winds of a few possible types of
relatively massive stars: asymptotic giant branch (AGB) stars with masses $4 \la
\cM/M_{\odot} \la 8$ \citep[e.g.,][]{vent+01}, or O-- or B-type stars that are
either rotating rapidly \citep{decr+07} or in interacting binary systems
\citep{demi+09}. 

A main problem in establishing the nature of the Na-O anticorrelation in GCs
has been finding out the timescale on which the abundance variations may have
occurred, which would have a direct impact on what type(s) of stars may have
been responsible. Significant renewed interest in this context was created by
the discovery of extended main sequence turn-offs (hereafter eMSTOs) in
massive intermediate-age (1\,--\,2 Gyr old) star clusters in the Magellanic Clouds
using high-quality Hubble Space Telescope \emph{(HST)} data 
\citep{macbro07,glat+08,milo+09,goud+09}. 
The typical star density distribution in such eMSTOs peaks in the bright blue
half and then decreases toward the faint red end, sometimes showing a
`hump' resembling a bimodal distribution \citep{mack+08a,milo+09,goud+11a}. 
Furthermore, several eMSTO clusters feature a faint extension to the
red clump of He-burning giants \citep{gira+09,rube+11,rube+13}. 

The nature of the eMSTO phenomenon is still highly debated. One common 
interpretation is the presence of an age spread of up to several $10^8$ yr
within these clusters \citep[see
also][]{rube+10,rube+11,rube+13,kell+11,gira+13,corr+14,goud+14,goud+15}.  
The main scenarios that have been suggested to provide the gas required for
extended star formation activity such age spreads are (1) mergers of (young)
clusters with giant molecular clouds \citep{bekmac09}, (2) accretion of
ambient gas by the clusters \citep{conspe11}, and (3) retention of 
stellar ejecta in the potential well of the young clusters
\citep{goud+11a,goud+14}. In this ``age spread'' scenario, the shape of the star
density distribution across the eMSTO mainly reflects the combined effects of
the histories of star formation and cluster dissolution due to strong cluster
expansion following the death of massive stars in the central regions
\citep[][hereafter G+14]{goud+14}. Support in favor of age spreads for
eMSTOs in intermediate-age clusters was provided by G+14 in the form of a
correlation between MSTO width and central escape velocity, which is a proxy for
the cluster's ability to retain and/or accrete gas during its early
evolution. G+14 also reported a strong correlation between the fractional
numbers of stars in the bluest region of the MSTO and those in the faint
extension of the red clump, as expected from an age spread.

However, non-detections of ongoing star formation in young massive clusters 
in nearby galaxies \citep{bast+13,cabr+14,cabr+16} have cast doubt on the
interpretation of eMSTOs as being due to age spreads (but see
\citealt{forbek17}). The leading alternative 
hypothesis on the nature of eMSTOs is that it is due to a spread of stellar
rotation rates. This idea was originally put forward by \citet{basdem09} who
argued that rotation lowers the luminosity and effective temperature at the
stellar surface (especially when viewed equator-on), which could cause an eMSTO
similar to those observed. However, more recent studies revealed a stronger
(and opposite) effect of stellar rotation, namely a longer main sequence (MS)
lifetime due to internal mixing \citep{gira+11,ekst+12,geor+13}, thus
mimicking a younger age (i.e., a brightening and blueing of the star at a
given mass).  

Recently, eMSTOs have also been detected in high-quality \emph{HST} data of much
younger clusters in the Large Magellanic Cloud (LMC) with ages of
$\sim$\,100\,--\,300 Myr
\citep{milo+15,milo+16,milo+17,bast+16,corr+15,corr+17}. These observations
yielded several pieces of evidence in support of the rotation hypothesis. One
such piece was the discovery of broadened or split main sequences which are
predicted to occur if a significant fraction of stars has rotation rates
$\Omega$ in excess of $\sim$\,80\% of the critical rate ($\Omega_{\rm C}$)
according to the Geneva {\sc syclist} isochrone models of \citet{geor+14},
whereas it cannot easily be explained by age spreads
\citep[see][]{dant+15,milo+16,corr+17}.   
Secondly, \citet{nied+15b} used the {\sc syclist} models and found that the
presence of a distribution of rotation rates in clusters with ages up to
$\sim$\,1 Gyr causes an eMSTO whose extent, if interpreted as an age spread, is
proportional to the cluster age, which is generally consistent with observations
(this finding will be further assessed in Sect.~\ref{s:spreads} below).  
Finally, narrow-band imaging studies of some young massive clusters in the LMC 
reveal that several stars in their MSTOs are strong H$\alpha$ 
emitters and thought to constitute equator-on Be stars that are known to be rapidly
rotating at $\OOc \ga 0.5$ \citep{bast+17,corr+17}. 

While these findings render it very likely that rotation is part of the solution
of the nature of eMSTOs, there are relevant indications that other effects
are at play as well. For example, the distribution of stars across the MSTO
of several young massive clusters with eMSTOs are \emph{not} consistent with a
coeval population of stars encompassing a range of rotation rates. Specifically,
the number of stars on the ``red'' side of the MSTO is significantly higher than
that expected if the ``blue'' side of the MSTO constitutes the bulk of the
rotating stars in a coeval population of stars, hinting at the presence of an
age spread in addition to a range of rotation rates
\citep{milo+15,milo+16,milo+17,corr+15,corr+17}. Furthermore, the relatively 
low-mass cluster NGC\,1844 with an age of $\sim$\,150 Myr features a broadened
MS as expected in the presence of a significant range of rotation rates, but
does \emph{not} exhibit an eMSTO even in high-quality \emph{HST} data 
\citep{milo+13}, in contrast to the massive 100 Myr old cluster
NGC\,1850 which does feature a clear eMSTO \citep{bast+16,corr+17}. 

In this paper, we present a re-analysis of the expected effects of spreads 
of stellar rotation rates within clusters on their eMSTO morphology as a
function of age, using the {\sc syclist} models, and we compare these effects
with observations to clarify and discuss some relevant shortcomings of the
rotation scenario.  
In Section~\ref{s:spreads}, we focus on measuring the extents of eMSTOs in
simulated cluster color-magnitude diagrams (CMDs) in the same way as done in
practice by observers. In Section~\ref{s:compare}, we compare {\sc syclist}
model predictions in detail with \emph{HST} observations of two massive
1-Gyr-old eMSTO clusters, near the upper end of the range of ages for which the
{\sc syclist} models can be used without uncertain extrapolations of the
effects of rotation to the photometric properties of stars. 
We discuss our results in the context of the rotation and age spread scenarios
in Section~\ref{s:disc} and we state our main conclusions in
Section~\ref{s:conc}.

\section{Evaluating Pseudo-Age Spreads due to Stellar Rotation} 
\label{s:spreads} 

Using the Geneva {\sc syclist} isochrone models, \citet{nied+15b} found
that the presence of a distribution of rotation rates in clusters with
ages up to $\sim$\,1 Gyr causes an eMSTO whose extent, if interpreted as
an age spread, is proportional to the cluster age. 
As shown in Figures 1 \& 2 of \citet{nied+15b}, this proportionality appears
when evaluating the difference in $M_V$ near the MSTO (hereafter
$\Delta\,M_{V,\,\rm MSTO}$) between isochrones of age $t_0$ with $\OOc$ =
0.0 and $\OOc = 0.5$\footnote{The maximum amount of brightening associated
  with stellar rotation in the {\sc syclist} models is reached at $\OOc = 0.5$.}
and then finding the age $t_1$ for which the same value of $\Delta\,M_{V,\,\rm
  MSTO}$ is found when comparing non-rotating isochrones of ages $t_0$ and
$t_1$. As such, they suggested that the age spreads found in the literature may
actually be due to ranges of stellar rotation rates rather than ranges of
stellar ages.

\begin{figure*}[tbhp]
\centerline{\includegraphics[width=0.85\textwidth]{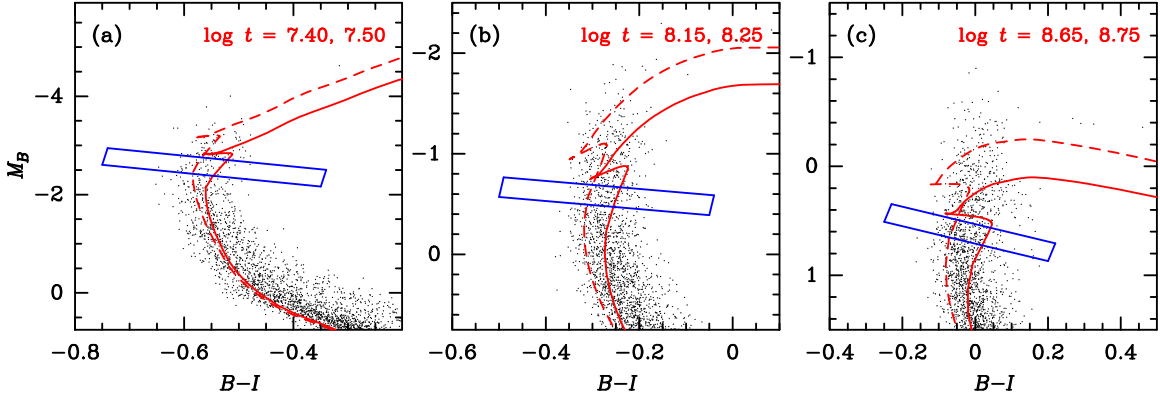}}
\caption{$M_B$ vs.\ $B\!-\!I$ CMDs of {\sc syclist} cluster simulations of ages
  $\log\, (t/{\rm yr})$ = 7.50, 8.25, and 8.75 (in panels a, b, and c,
  respectively), using the distribution of rotation rates by
  \citet{huan+10}. Note that the axis ranges are different for each panel.  
  {\sc syclist} isochrones for non-rotating stars are overplotted for ages
  shown in the legend of each panel (in each case, the isochrone for the first
  age mentioned is plotted with dashed lines). Parallelograms used to determine
  pseudo-age distributions are shown with blue solid lines. See discussion in 
  Section~\ref{s:spreads}. 
 }
\label{f:clustersims}
\end{figure*}

However, one potential flaw in the reasoning of \citet{nied+15b} is that
observational studies of intermediate-age clusters did \emph{not} determine
(pseudo-)age spreads in terms of \emph{vertical} extents of their MSTOs on the
CMD. Instead, most such studies measured the extent of the MSTO along a
direction approximately  perpendicular to the isochrones in the MSTO region,
which is closer to  \emph{horizontal} on the CMD (i.e., parallel with color)
than to vertical \citep[see,
e.g.,][]{milo+09,goud+11b,goud+14,piabas16,corr+14,corr+15,corr+17}. To 
evaluate the degree to which this changes the results of \citet{nied+15b}, we
conduct Monte Carlo simulations using the {\sc syclist} cluster models 
through the website described in 
\citet{geor+14}\footnote{\url{http://obswww.unige.ch/Recherche/evoldb/index}.}.
The differences of our method relative to that of \citeauthor{nied+15b}
constitute not only the use of a more realistic definition of pseudo-ages, but
also the inclusion of orientation and population effects, and observational
errors. These effects cannot be considered with a simple approach that
compares two isochrones.  

For these simulations, we adopt the following choices: 
\emph{(i)} the number of stars is 10,000; \emph{(ii)} a random distribution of
inclination angles with respect to the line of sight; \emph{(iii)} gravity
darkening as described by \citet{esprie11}; \emph{(iv)} limb darkening as
described by \citet{clar00}; \emph{(v)} the distribution of $\OOc$ as
described by \citet[][hereafter HGM10]{huan+10}.
We selected this distribution because
it has the highest percentage of very rapidly rotating 
  stars (e.g., $\OOc \ga 0.9$) among the ones available through the {\sc
    syclist} web interface (cf.\ Section~\ref{s:compare}); \emph{(vi)} a
  binary fraction of 0.3, similar to those found for young and
  intermediate-age clusters by many \emph{Hubble Space Telescope (HST)}
  studies  
\citep[e.g.,][]{milo+09,goud+11b,goud+14,milo+15,milo+16,corr+14,corr+15}\footnote{The  
  results of these simulations are insensitive to the chosen binary fraction
  since the crosscuts through the MSTO are done in a region where the binary
  sequence joins that of single stars.}, and \emph{(vii)} a
metallicity of $Z = 0.006$. To account for photometric
noise at a level that is typical for \emph{HST} (ACS or WFC3) data of
star clusters in the LMC, we use the extensive artificial star tests
by \citet{goud+11b} and G+14 to calculate the average relations between magnitude
and magnitude error ($\sigma$) among their 16 intermediate-age clusters in the 
LMC at the core radius of the clusters in question. This is done for
filters close to Johnson $B$ (i.e., \emph{F435W} for ACS data and
\emph{F475W} for WFC3 data) and Cousins $I$ (i.e., \emph{F814W} for
ACS and WFC3 data). Magnitude errors are added to the simulated
magnitudes (assuming $m\!-\!M$ = 18.50 at the distance of the LMC) by drawing
random numbers from a Gaussian distribution using values of $\sigma$ from the
relation mentioned above. We assume that the photometric errors derived this 
way constitute an adequate estimate for the great majority of recent
high-quality observations of star clusters in the LMC with the ACS and WFC3
cameras aboard \emph{HST}. This includes all data shown below in
Figure~\ref{f:dAge_vs_Age} with the possible exception of those of 
\citet{nied+15a} who used somewhat lower quality data from the WFPC2 camera. 

To determine ``pseudo-age spreads'' for these {\sc syclist} cluster simulations
in a way similar to that done for the majority of studies of eMSTO clusters
using \emph{HST} photometry, we measure the positions of stars along the long
axis of a parallelogram drawn across the eMSTO in a $M_B$ vs.\ $B\!-\!I$ CMD
in a direction approximately perpendicular to the non-rotating isochrones
and hence approximately parallel to age for non-rotating stars \citep[for a
full description of such pseudo-age distributions, see][]{goud+11a}. The long
axis of such parallelograms, which are illustrated in 
Figure~\ref{f:clustersims} for three examples of simulated clusters with ages in
the range $10^{7.5}$\,--\,$10^{9.0}$ yr, are translated to pseudo-ages of
non-rotating models in the following way. First we produce a grid of 
non-rotating {\sc syclist} isochrones covering the extent of the eMSTO, and then 
use a least-squares fit to determine the relation between age and the
coordinate along the long axis of the parallelogram; this relation is then
used to convert the position of any simulated star --rotating or not-- into a
pseudo-age. Representative pseudo-age spreads of the simulated 
clusters are then determined by measuring the FWHM of their pseudo-age
distributions.  

\begin{figure}[tbhp]
\centerline{\includegraphics[width=7.5cm]{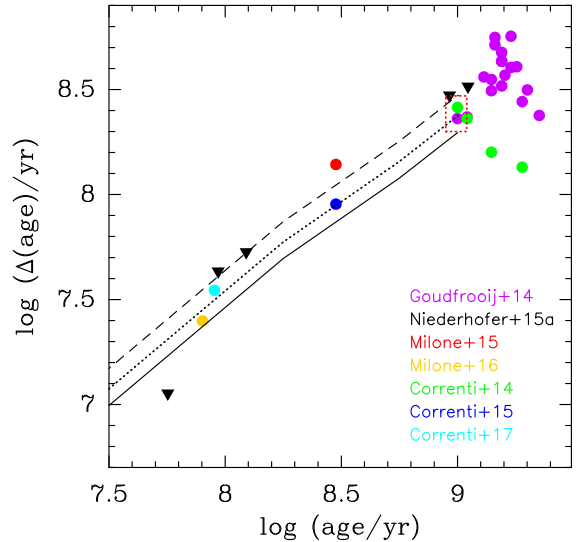}}
\caption{Predicted versus observed relations between pseudo-age spread in eMSTO
  clusters in the Magellanic Clouds and their age. The solid black line
  represents the prediction if a wide range of stellar rotation rates is
  interpreted as an age spread according to the {\sc syclist} models, based on
  FWHM values of pseudo-age distributions of simulated clusters as described in
  the text. For comparison, the dotted and dashed black lines
    show the same for age spreads that are larger than those indicated by the
    solid line by 20\% and 50\%, respectively. 
  The filled symbols represent observed pseudo-age spreads of young
  and intermediate-age clusters. Colors indicate the respective studies as
  shown in the legend. The data points for NGC\,1987 and NGC\,2249 are
  surrounded by a dotted box. Note that the {\sc syclist} model
    predictions are only available for ages $\la 10^9$ yr. 
 }
\label{f:dAge_vs_Age}
\end{figure}

The results are shown as a solid black line in Figure~\ref{f:dAge_vs_Age},
which compares these results to the FWHM pseudo-age spreads measured for real 
clusters in the same age range by various authors using 
different methods\footnote{As such, we multiplied the standard
  deviation values of \citet{nied+15a} by a factor $\sqrt{8\, \ln 2}$ prior
  to plotting.}
(G+14, \citealt{nied+15a,milo+15,milo+16,corr+14,corr+15,corr+17}).
The overall picture is similar to that described by \citet{nied+15b} in the
sense that the inferred pseudo-age spreads are proportional to the actual ages
of the clusters. However, the \emph{absolute} pseudo-age spreads inferred for
a given age when measured across the width of the eMSTO as mentioned
above are smaller than those evaluated by \citet{nied+15b}. The net result is
that the pseudo-age spreads measured in the literature for many clusters in
the age range of $10^{7.5}$\,--\,$10^{9.0}$ yr are $\approx 20-50$\% larger
than can be accounted for by a coeval population of stars encompassing a large
range of  rotation rates according to the {\sc syclist} models. This result
will be discussed further below.   

\section{Comparison with Observations of NGC\,1987 and NGC\,2249} 
\label{s:compare}

To illustrate the impact of the analysis shown in the previous section to the
nature of eMSTO's in intermediate-age clusters, we compare the {\sc syclist}
isochrone models with high-quality \emph{HST} 
photometry of NGC\,1987 and NGC\,2249, two eMSTO clusters in the LMC with ages
of $\sim$\,1 Gyr. These clusters were selected for this purpose for two
  main reasons. Firstly, 
their age is very close to the oldest age for which the relevant part of the
MSTO still consists of stars with $\cM \geq 1.7\;M_{\odot}$ at $Z = 0.006$, so
that the {\sc syclist} models can still be used without having to rely on
extrapolations of rotational correction factors. The latter are needed at
$\cM < 1.7\;M_{\odot}$, where stellar atmospheres undergo complex transitions
that have a significant (but difficult to quantify) impact on stellar rotation:
the transition from a convective to a radiative core and the transition from a
radiative to a convective envelope. Secondly, the color spread
  covered by eMSTOs is larger for older ages, allowing their morphology
  to be studied in more detail than in younger clusters. 

The \emph{HST/ACS} photometry of NGC\,1987 was described before
by \citet[][\emph{F435W} and \emph{F814W} filters]{goud+11b}, while
the \emph{HST/WFC3} photometry of NGC\,2249 was described by
\citet[][\emph{F438W} and \emph{F814W} filters]{corr+14}. 
The data files are available at the STScI MAST Archive 
\href{https://doi.org/10.17909/T9Q606}{\color{MyBlue}(10.17909/T9Q606)}. 
To compare the {\sc syclist} isochrone models with 
these \emph{HST} datasets, we  
proceed as follows. For the {\sc syclist} isochrones of non-rotating stars, we
first integrate the specific intensities of the ATLAS9 model atmospheres of
\citet{caskur03} over the projected stellar surface, and convolve the emerging 
model spectra with the \emph{HST} filter transmission curves available through
{\sc synphot} \citep{laid+08} to calculate absolute magnitudes in the
\emph{HST} passbands. 
For the resulting isochrones, we determine linear relations
between $I$ and \emph{ACS/F814W}, $I$ and \emph{WFC3/F814W}, $B\!-\!I$ and
\emph{F435W\,--\,F814W}, and $B\!-\!I$ and \emph{F438W\,--\,F814W} for stars
within the intervals of $\log\,L$ and $\log\,T_{\rm eff}$ that cover the full
extent of the MSTO region in the CMD. These relations were then applied to the
$B$ and $I$ magnitudes in the {\sc syclist} isochrones (for any value of
$\OOc$) to derive magnitudes in the \emph{HST} passbands.  

\begin{figure*}[tbhp]
\centerline{\includegraphics[width=13cm]{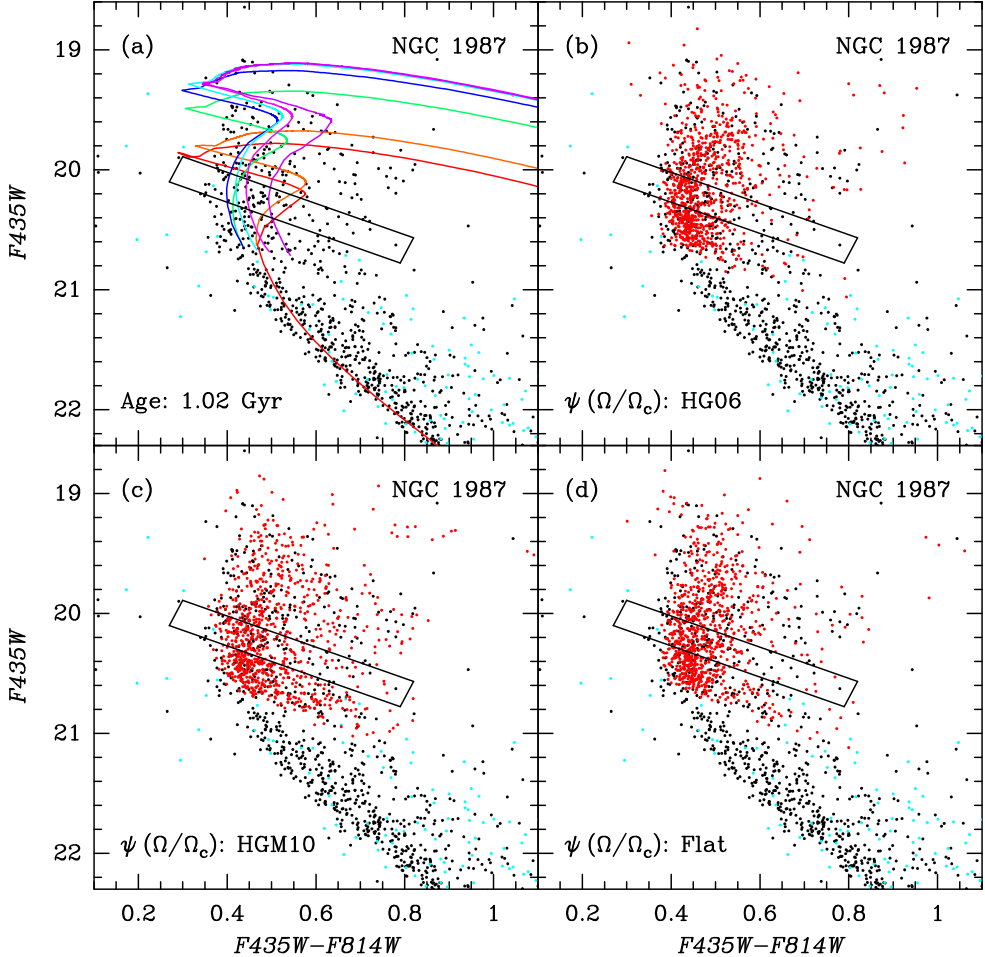}}
\caption{Upper MS and MSTO regions of CMD of NGC\,1987. In all panels, the black
  dots represent stars within the core radius of NGC 1987 while cyan dots
  represent stars in a background region of the image of NGC 1987 with the same
  area as that shown by the black dots. The parallelogram used to derive
  pseudo-age distributions (see Sect.\ \ref{s:compare}) is also shown in all panels. 
  \emph{Panel (a)}: the overplotted lines represent {\sc syclist} isochrones for
  an age of 1.02 Gyr and the following values of $\OOc$: 0.00 (red line), 0.10
  (orange line), 0.30 (green line), 0.50 (blue line), 0.70 (cyan line), 0.80 
  (magenta line), and 0.90 (purple line). 
  \emph{Panel (b)}: the red dots represent a simulated cluster using the {\sc
    syclist} models for an age of 1.02 Gyr and the $\OOc$ distribution of HG06.
  \emph{Panel (c)}: same as panel (b) but now for the $\OOc$ distribution of HGM10.
  \emph{Panel (d)}: same as panel (b) but now for a flat distribution of $\OOc$.
 }
\label{f:CMD1987}
\end{figure*}

\begin{figure*}[tbhp]
\centerline{\includegraphics[width=13cm]{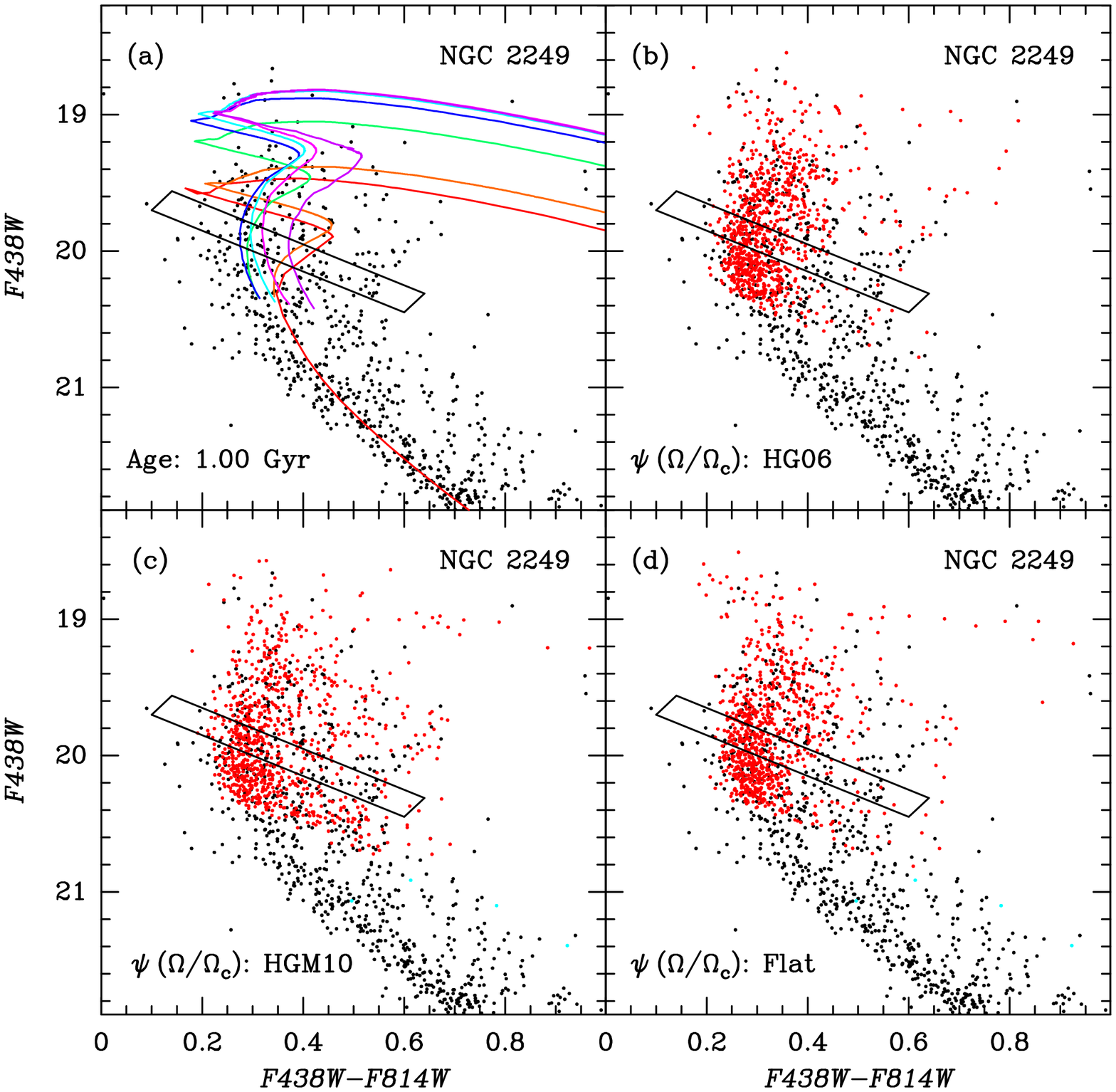}}
\caption{Same as Figure~\ref{f:CMD1987}, but now for NGC 2249. The {\sc syclist}
  isochrones are for an age of 1.00 Gyr in this case. 
 }
\label{f:CMD2249}
\end{figure*}

CMDs of NGC\,1987 and NGC\,2249 are presented in Figures~\ref{f:CMD1987} and
\ref{f:CMD2249}. In order to keep the
contamination of the MSTO by LMC field stars to negligible levels, we select
stars in NGC\,1987 within its core radius, while for NGC\,2249, we select
stars within its effective radius. For each cluster, we use the
non-rotating Geneva isochrones for $Z = 0.006$ from \citet{mowl+12} to find
combinations of age, distance modulus ($\mM$), and reddening ($A_V$) to
provide a good fit to the cluster's MS and the middle of its MSTO. For the
same combinations of age, $\mM$, and $A_V$, we then overplot the {\sc syclist}
isochrones from \citet{geor+13} for various values of $\OOc$ to illustrate the
effect of the ``apparent rejuvenation'' caused by a distribution of rotation
rates to the extent of the MSTO\footnote{For a given age, $Z$, 
  $\cM/M_{\odot}$, and filter passband, the \citet{geor+13} isochrones for $\OOc
  = 0.0$ are offset from the \citet{mowl+12} ones by a few hundredths of a
  magnitude due to differences in the input physics. This offset was applied to
  the \citet{geor+13} isochrones here.}. After comparing this set of isochrones
to the data, the age is adjusted slightly until we obtain a good fit to both the
top and the left edge of the MSTO (by the isochrones with $\OOc \geq 0.5$). The
resulting sets of isochrones are shown in panel (a) of Figures~\ref{f:CMD1987}
and \ref{f:CMD2249}.  

As mentioned in the previous Section, the vertical extent of eMSTOs is well
fitted by a set of single-age isochrones that represent a wide distribution of
stellar rotation rates, as also emphasized by \citet{nied+15b}. This is the case for
both NGC\,1987 and NGC\,2249. However, it is also apparent that the eMSTO in both
clusters extends to significantly redder colors than those covered by the set of
isochrones representing different rotation rates. 
To test whether this discrepancy can be accounted for by the cumulative effects
of different inclination angles of rotating stars with respect to the line of
sight as well as gravity and limb darkening, we conduct Monte Carlo simulations
using the {\sc syclist} cluster models at the ages of NGC\,1987 and NGC\,2249
as described in Section~\ref{s:spreads} above. 
Since these cluster models only involve stars with $\cM \geq 1.7\;M_{\odot}$, the
number of stars in these particular simulations was chosen to be equal to 
the observed number of cluster stars in the MSTO and post-MS regions on the
CMD, rounded to the closest hundred.  
Since there is no a priori knowledge of the distribution of $\OOc$ in these
clusters, we create separate simulations for all such distributions available
in the {\sc syclist} models, namely those of \citet[][hereafter
HG06]{huagie06}, \citet[][hereafter HGM10]{huan+10}, and a flat
distribution. In addition, we create simulations for a bimodal 
distribution in which 2/3 of the stars rotate very rapidly ($\OOc$ = 0.9) while
the remaining stars have negligible rotation ($\OOc$ = 0.0). This distribution
was shown by \citet{dant+15} to provide a good fit to the observed morphology
of the split MS in the young massive LMC cluster NGC\,1856 \citep[see
  also][]{bast+17,corr+17}.
 While intermediate-age clusters like NGC~1987 and NGC~2249 do not exhibit
  split MSes, this is likely due to their MS stars having masses
  $\cM < 1.7\;M_{\odot}$ which feature magnetic braking as well as less
  effective mixing of envelope material into the core than do stars with
  $\cM > 1.7\;M_{\odot}$ (see Sect.\ \ref{sub:rotscenario}.2;
  \citealt{brahua15}). As such, the absence of split MSes in these clusters
  does not necessarily equate to the absence of a bimodal distribution of
  $\OOc$.    

Panels (b)\,--\,(d) in Figures~\ref{f:CMD1987} and \ref{f:CMD2249} illustrate
the MSTO morphologies produced by the {\sc syclist} cluster models described
above, for the three different $\OOc$ distributions.  Note that the effect of
random inclinations in conjunction with gravity and limb darkening can indeed
have a significant impact on the resulting MSTO morphology of $\sim$\,1-Gyr old
clusters.  Specifically, the red end of the eMSTO can be populated by equator-on
rapid rotators, which is especially apparent for the $\OOc$ distribution of
HGM10. As such, the full \emph{extent} of the eMSTOs of these clusters 
can in principle be covered by a coeval population of stars with varying
rotation rates. This is consistent with the findings of \citet{brahua15}. 

\begin{figure}[tbh]
\centerline{\includegraphics[width=8cm]{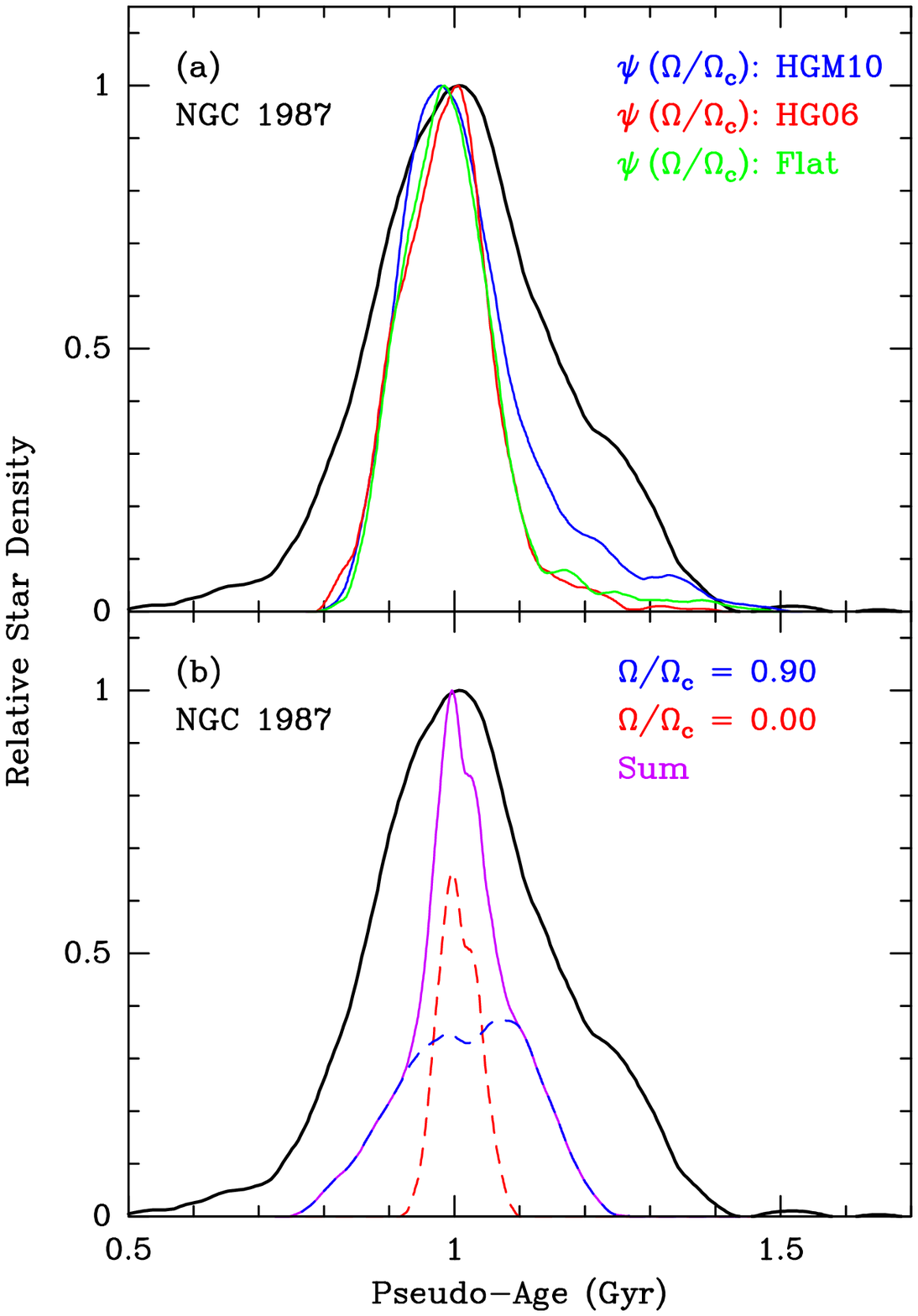}}
\caption{\emph{Panel (a)}: The thick black solid line represents the pseudo-age
  distribution of NGC\,1987, while the other solid lines represent the
  pseudo-age distributions {\sc syclist} simulations of a cluster with an age of
  1.02 Gyr and different distributions of $\OOc$, as indicated in the
  legend. 
  \emph{Panel (b)}: Similar to panel (a), but now the blue and red dashed
  lines represent the pseudo-age distributions of stars with $\OOc$ = 0.9 and
  $\OOc$ = 0.0 in the bimodal distribution of \citet{dant+15}, respectively. The
  purple line represents the sum of the two latter distributions. See discussion
  in Sect.\ \ref{s:compare}. 
 }
\label{f:crossMSTO_1987}
\end{figure}

\begin{figure}[tbh]
\centerline{\includegraphics[width=8cm]{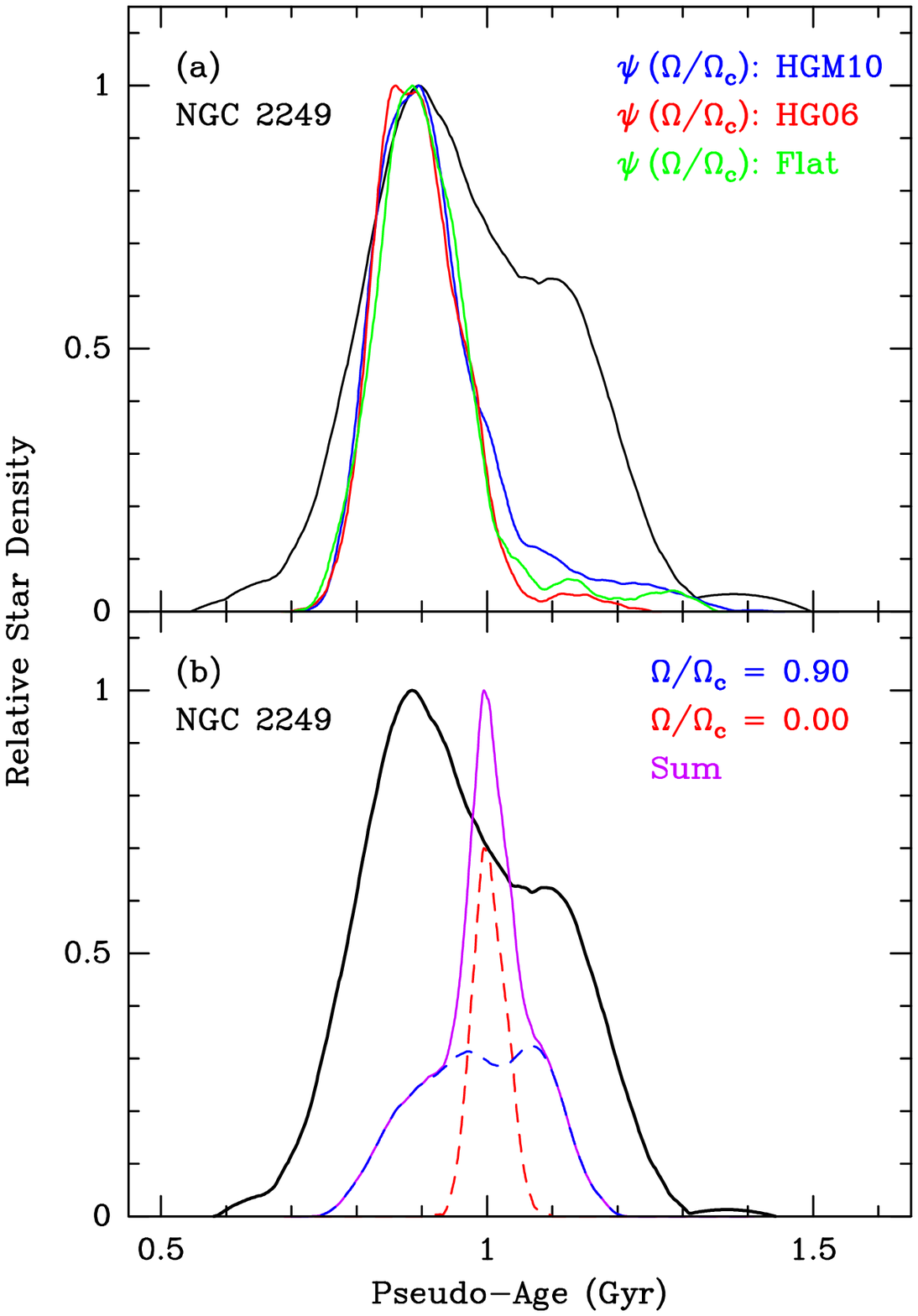}}
\caption{Same as Figure~\ref{f:crossMSTO_1987}, but now for NGC\,2249. 
 }
\label{f:crossMSTO_2249}
\end{figure}

However, the \emph{distribution} of stars across the eMSTO in these clusters
turns out to be significantly different from that predicted by the {\sc
  syclist} models 
under the assumption of random inclination angles. 
To illustrate this, we derive pseudo-age distributions of
the two clusters, as well as their respective {\sc syclist} cluster
simulations\footnote{The pseudo-age distributions of the {\sc syclist}
  simulations shown in Figures~\ref{f:crossMSTO_1987} and
  \ref{f:crossMSTO_2249} constitute averages of 15 simulations each.}, in the
way described in Section~\ref{s:spreads}. These are shown in
Figures~\ref{f:crossMSTO_1987} and \ref{f:crossMSTO_2249} using solid lines of 
different colors. Note that the red (or ``old'') half of the pseudo-age
distributions of both clusters contain significantly more stars (relative to the
peak in the blue or ``young'' half) than those in any of the single-age
{\sc syclist} simulations, especially in the case of NGC\,2249. 

\section{Discussion}
\label{s:disc}

\subsection{Constraints on The Rotation Scenario}
\label{sub:rotscenario}

\subsubsection{Constraints from NGC 1987 and NGC 2249}

\begin{figure}[tbhp]
\centerline{\includegraphics[width=8cm]{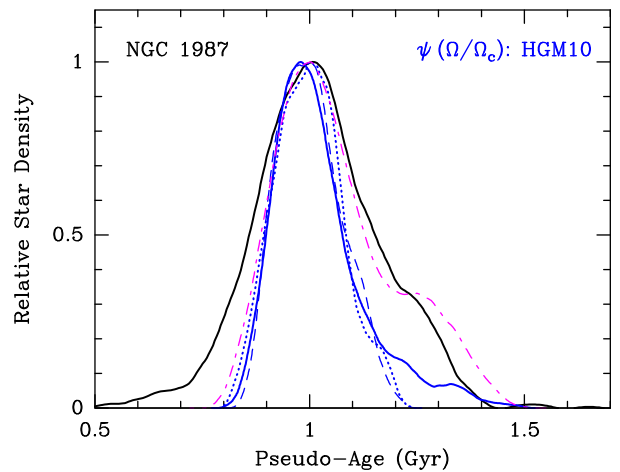}}
\caption{Similar to Figure~\ref{f:crossMSTO_1987}, except that the
  only {\sc syclist} simulation shown is the one using the HGM10
  distribution of $\OOc$ (solid blue line). The dashed and dotted blue
  lines represent the same {\sc syclist} simulation, but now ignoring the
  effects of gravity and limb darkening (dashed blue line) and
  selecting only stars with viewing angles between 40\degr\ and
  50\degr\ (dotted blue line), respectively, 
  while the magenta dot-dashed line shows the same {\sc
      syclist} simulation after selecting only stars with viewing angles
    between 80\degr\ and 90\degr.  
  See discussion in Sect.\ \ref{sub:rotscenario}.1. 
 }
\label{f:crossMSTO_1987_rot}
\end{figure}

\begin{figure}[tbhp]
\centerline{\includegraphics[width=8cm]{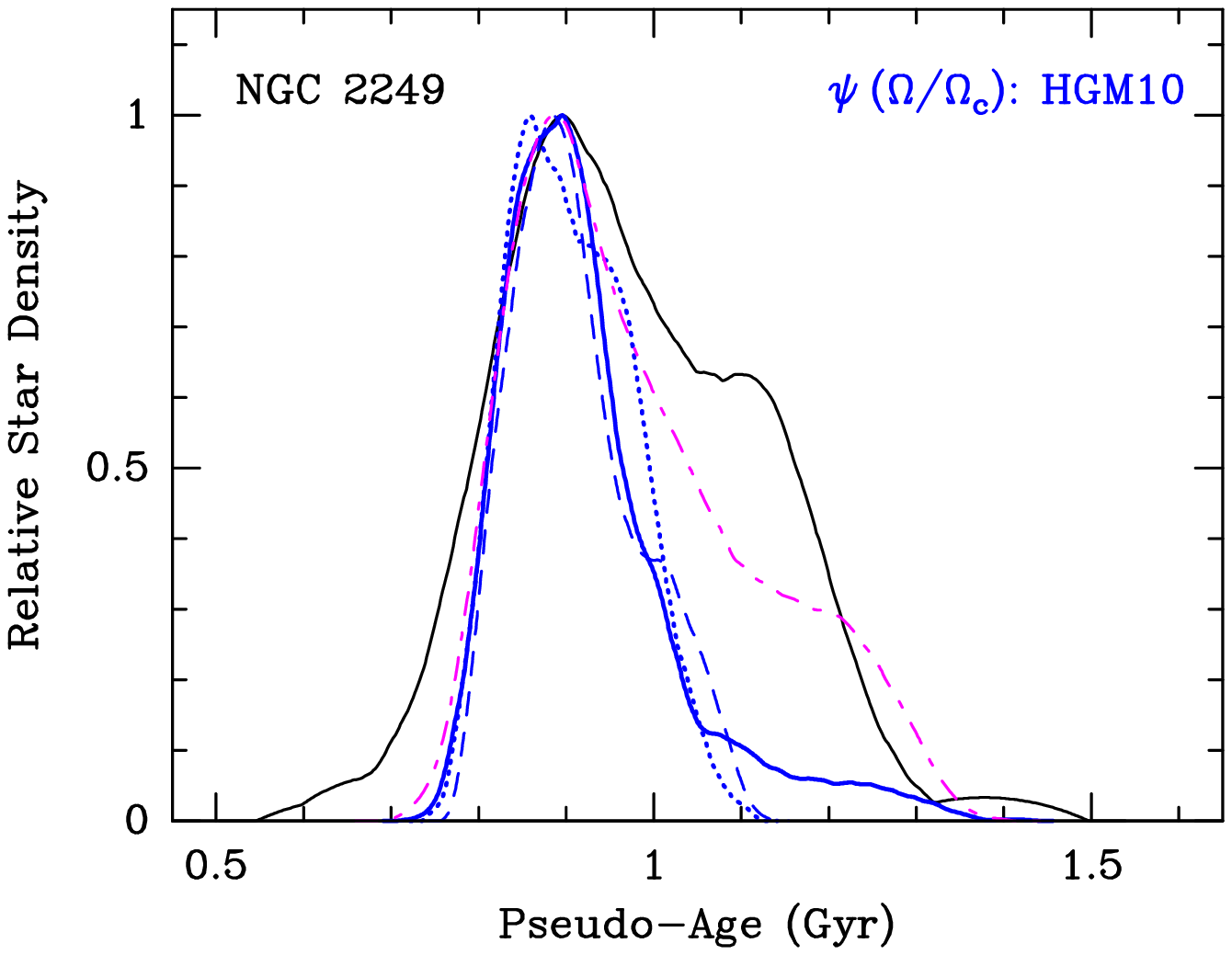}}
\caption{Same as Figure~\ref{f:crossMSTO_1987_rot}, but now for NGC 2249. 
 }
\label{f:crossMSTO_2249_rot}
\end{figure}

\noindent
Attempting to put the discrepancy between the observed pseudo-age distributions
in NGC\,1987 and NGC\,2249 and the various single-age {\sc syclist} simulations
in context, we first consider the relative impacts of \emph{(i)} gravity and
limb darkening, and \emph{(ii)} the distribution of inclination angles of 
rotating stars with respect to the line of sight. To address point \emph{(i)},
we use the same {\sc syclist} cluster simulations and use the values of log $L$
and log $T_{\rm eff}$ that do \emph{not} include the effects of gravity and limb
darkening. To address point \emph{(ii)}, we only select stars in the 
simulations that have moderate viewing angles $i$ of $40^{\circ} \leq
i \leq 50\degr$. The pseudo-age distributions of those two cases are
shown for the HGM10 distribution of $\OOc$ in
Figures~\ref{f:crossMSTO_1987_rot} and 
\ref{f:crossMSTO_2249_rot} as dashed and dotted lines,
respectively. Note that the two effects have very similar impacts on 
the resulting pseudo-age distribution. If we accept the magnitude of
gravity and limb darkening effects as implemented in the {\sc syclist}
models, it follows that the only way to make these models fit the
observed pseudo-age distributions of NGC\,1987 and NGC\,2249 in an
adequate manner is to require not only a distribution of $\OOc$ that
includes a substantial fraction of very rapid rotators ($\OOc \ga 0.90$),
but also \emph{a distribution of viewing angles that is strongly skewed toward
  equator-on configurations}.  
This is illustrated by the magenta dot-dashed lines in
Figures~\ref{f:crossMSTO_1987_rot} and \ref{f:crossMSTO_2249_rot}, which show
the pseudo-age distributions of stars in the simulations that have viewing
angles close to equator-on: $80\degr \leq i \leq 90\degr$. These curves 
provide a significantly improved fit to the observed pseudo-age distributions, 
especially for the case of NGC\,1987. While it may seem intuitively unlikely
that the viewing angles of rotating stars within a star cluster would only
cover such a small range, a recent asteroseismology study of two Galactic open
clusters did recently find the rotation axes of $\sim$\,75\% of the observed
red giant stars in each of those clusters to be closely aligned with each 
other \citep{cors+17}, suggesting that the kinetic energy of proto-cluster
molecular clouds can be dominated by rotation (rather than turbulence) which
can in turn be transferred to its constituent stars during the star formation
process in an efficient manner.  

However, the fact that one would need the viewing angles to be close to
equator-on for the rotation scenario to work for both NGC\,1987 and NGC\,2249
seems suspect. The probability that an intrinsically random viewing angle is 
$80\degr \leq i \leq 90\degr$ for two clusters is $\sim$\,0.01; taking into
account that the real distribution of $\OOc$ may be different from those
supported in the {\sc syclist} models, the actual probability may be slightly
higher than that but should not exceed a few per cent. Furthermore, the
pseudo-age distributions of NGC\,1987 and NGC\,2249 are \emph{narrower than
  those of most other intermediate-age clusters in the Magellanic Clouds},
rendering it unlikely that rotation alone
(as implemented in the current {\sc syclist} models) can explain the
eMSTO widths of all such clusters.  This is discussed further in
Section~\ref{sub:rotscenario}.2.

Comparing the observed pseudo-age distributions of NGC 1987 and NGC 2249 with 
the simulations that represent the bimodal distribution of $\OOc$ advocated by
\citeauthor{dant+15}\ (\citeyear{dant+15}; see Figures~\ref{f:crossMSTO_1987}b and
\ref{f:crossMSTO_2249}b), we note that the latter simulations produce star
densities that are significantly lower than the observed ones at both low and high
pseudo-age values (i.e., the ``top left'' and ``bottom right'' ends of the
eMSTO, respectively). As shown by panel (a) of Figures~\ref{f:CMD1987} and
\ref{f:CMD2249}, stars in the top left of the eMSTO correspond mainly to
stars with $0.2 \la \OOc \la 0.8$ in the rotation scenario. As such, the lack
of stars in this region in the simulation representing the \citet{dant+15}
distribution suggests that the distribution of $\OOc$ in these clusters is more
uniform than purely bimodal ($\OOc$ = 0.0 and 0.9) in the context of the
stellar rotation scenario. Conversely, the relative lack of stars in the bottom
right of the eMSTO can only be explained in the rotation scenario by assuming an
very high proportion of equator-on viewing angles, as discussed above.

\subsubsection{Constraints from clusters with ages 1.4\,--\,2.0 Gyr}

While the general correlation between MSTO extent and age among eMSTO clusters
first shown by \citet{nied+15b} is quite well established among clusters with ages $\la 1$
Gyr (see also Section~\ref{s:spreads}), there are some features of eMSTOs
among clusters in the Magellanic Clouds with somewhat older ages
($\sim$\,1.4\,--\,2.0 Gyr) that seem to challenge the rotation
hypothesis. These are described below. 

One such feature is that relative to the two 1-Gyr-old clusters studied in
this paper (NGC\,1987 and NGC\,2249), the average width of eMSTOs increases
with cluster age, peaking around an age of 1.5\,--\,1.6 Gyr, followed by a decrease
\citep[see, e.g., G14;][]{nied+15b,piabas16}\footnote{The age where the peak
  average width of the eMSTOs occurs is $\approx$\,2.0 Gyr in the study of
  \citet{piabas16}.}. In the context of the {\sc syclist} models, one would
instead expect the width of eMSTOs to decrease at ages beyond $\sim$\,1.0 Gyr at the
metallicities of the LMC and SMC \citep[see also][]{brahua15}, since the MSTO 
at older ages is populated by stars with $\cM/M_{\odot} < 1.7$ at those
metallicities. Such stars have convective envelopes whose angular momentum is
decreased considerably by a magnetized wind. Furthermore, the cores of such
lower-mass stars are less convective than those with $\cM/M_{\odot} > 1.7$,
thus providing less mixing at their core boundaries and allowing less
additional ``rejuvenating'' Hydrogen fuel from the outer regions to burn in
their cores\footnote{These complex effects were not accounted for by
  \citet{brahua15} for stars with $\cM/M_{\odot} < 1.7$. They implicitly
  assumed that the effects of rotation do not change between 1.45 and 1.7
  $M_{\odot}$, thus overestimating the effect of rotation on the widths of
  eMSTOs of Magellanic Cloud clusters at ages $\ga 1.0$ Gyr.}.
 While the current {\sc syclist} models do not allow one to test
  this directly, 
the low pseudo-age spreads found for some of the lower-mass LMC clusters with
ages $>$ 1.0 Gyr studied by \citet{corr+14} are consistent with the notion
that the impact of rotation to the extent of MSTOs indeed diminishes at
stellar masses $\cM/M_{\odot} < 1.7$ (see green circles in
Figure~\ref{f:dAge_vs_Age}).   The increase of the average width of eMSTOs
beyond an age of 1.0 Gyr therefore suggests that some property other than
stellar rotation is partly responsible for the eMSTO phenomenon, at least for
those older clusters.   

Another feature among eMSTOs of clusters in the narrow age interval of
1.4\,--\,1.7 Gyr is that their widths vary significantly among clusters. This
is especially clear in the sample of G+14 (see also
Figure~\ref{f:dAge_vs_Age} and \citealt{piabas16}), where the FWHM of the
pseudo-age distributions of clusters in this age interval ranges between
about 200 and 550 Myr and correlates with cluster mass and escape velocity
(G+14). Taken at face value, this seems inconsistent with the rotation
hypothesis as represented by the current {\sc syclist} models, 
since we showed above that the viewing angles already need to be close to
equator-on to explain the pseudo-age distributions of NGC\,1987 and NGC\,2249
(whose FWHMs are only $\la 250$ Myr) in the context of the rotation hypothesis. 
Conversely, this feature \emph{is} consistent with the ``age range''
hypothesis of G+14 in which the extent of the eMSTO reflects the period during
which the cluster is able to form stars from retained and/or accreted gas at
young ages.

\subsection{Constraints on Age Spreads}
\label{sub:agescenario}

Accepting the evidence for the presence of a wide range of rotation
velocities in star clusters from the observations of young massive clusters in
the LMC \citep{dant+15,milo+16,milo+17,bast+17,corr+17}, we now address 
implications of our results on the ``age spread'' scenario, i.e., the
idea that eMSTOs are mainly due to spreads in stellar age.  
  
\begin{figure}[tbhp]
\centerline{\includegraphics[width=8cm]{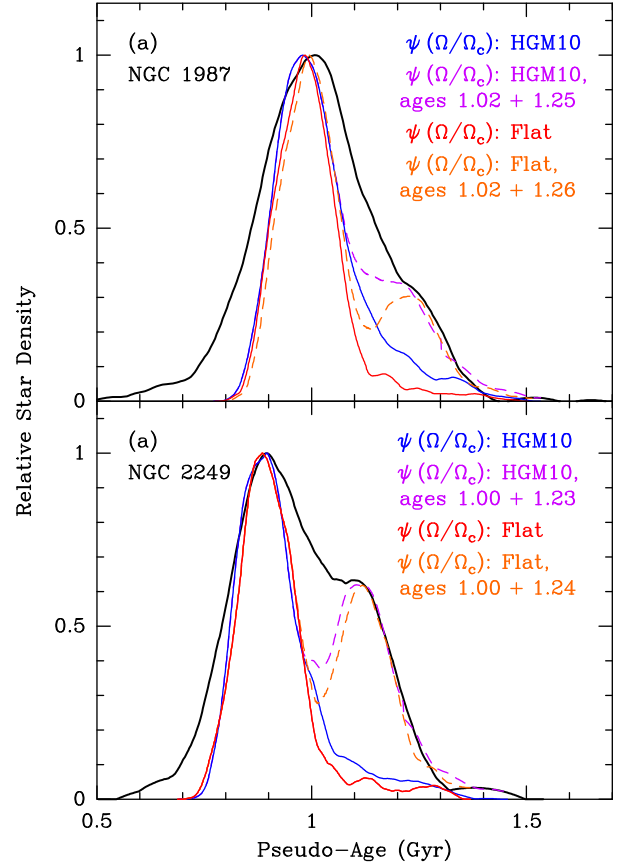}}
\caption{\emph{Panel (a)}: Same as panel (a) of Figure~\ref{f:crossMSTO_1987},
  to which we added dashed lines that represent the sum of two {\sc
    syclist} cluster simulations involving random viewing angles: the purple
  line represents the $\OOc$ distribution of HGM10 with ages of 1.02 and 1.25
  at relative weights of 1.0 and 0.20, respectively, and the orange line
  represents a flat $\OOc$ distribution with ages of 1.02 and 1.26 at
  relative weights of 1.0 and 0.27, respectively. 
  \emph{Panel (b)}: Same as panel (a) of Figure~\ref{f:crossMSTO_2249},
  to which we added dashed lines that represent the sum of two {\sc
    syclist} cluster simulations: the purple line represents the $\OOc$
  distribution of HGM10 with ages of 1.00 and 1.23 at relative weights of 1.0
  and 0.53, respectively, and the orange line represents a flat $\OOc$
  distribution with ages of 1.00 and 1.24 at relative weights of 1.0
  and 0.56, respectively. 
  See discussion in Sect.\ \ref{sub:agescenario}. 
 }
\label{f:crossMSTO_1987_2249_w_ages}
\end{figure}

Under the assumptions that \emph{(i)} the distribution of $\OOc$ is fairly
smooth (rather than showing discrete peaks), especially in the range $0.3 \la
\OOc \la 1.0$, and \emph{(ii)} the viewing angles of the stellar rotation axes
with respect to the line of sight are distributed randomly, the simulations
described in the previous sections agree with results of other recent studies in
that the presence of a range of stellar rotation rates does increase the extent
of MSTOs of simple stellar populations (SSPs). Since our previous
investigations of eMSTOs in intermediate-age 
clusters in the Magellanic Clouds used simulations with non-rotating isochrones
to quantify spreads of stellar age necessary to fit pseudo-age distributions
\citep{goud+09,goud+11a,goud+14}, one might thus expect that the real age spreads
should be smaller. To quantify this for NGC\,1987 and NGC\,2249, we
take the pseudo-age distributions of {\sc syclist} SSP simulations
shown in Figures~\ref{f:crossMSTO_1987}a and \ref{f:crossMSTO_2249}a
and add scaled versions of the same pseudo-age distributions for older
stellar ages until we obtain a good fit to the right-hand part of the
observed pseudo-age distributions of the cluster in
question\footnote{While there is a discrepancy between the observed
  and simulated pseudo-age distributions of NGC 1987 in the ``young'' wing as
  well, we do not put a significant weight on this in view of the uncertainty
  of the subtraction of field star contamination in this region of the
  CMD for this cluster (cf.\ Figure~\ref{f:CMD1987}a; 
  \citealt{goud+11b}).}. We use an age step of 0.01 Gyr in this context. 
 Note that this procedure implicitly assumes that the effects of rotation
  to the pseudo-age distribution do not diminish at ages $> 1.20$ Gyr, which
  is the oldest age for which the {\sc syclist} isochrones cover the
  parallelogram in Figures~\ref{f:CMD1987} and \ref{f:CMD2249}. 
The results are shown for the $\OOc$ distributions of HGM10 and a flat
distribution as the dashed curves in
Figure~\ref{f:crossMSTO_1987_2249_w_ages}.     

The good fit to the ``old'' edge of the observed pseudo-age distributions by
these simulations illustrates that the inability of the stellar rotation
scenario to fit the observed pseudo-age distributions of massive
intermediate-age star clusters such as NGC\,1987 and NGC\,2249 can be
effectively overcome by introducing a suitable distribution of stellar
ages. In the specific cases of NGC\,1987 and NGC\,2249, the best-fit 
age spreads are $235 \pm 10$ Myr and $235 \pm 10$ Myr, respectively, where the
quoted uncertainties reflect the differences found between the assumptions of
the $\OOc$ distribution of HGM10 and a flat distribution. These age spreads
are similar or somewhat ($\la 15$\%) smaller than those estimated from the FWHM
values of the observed pseudo-age distributions of NGC\,1987 and NGC\,2249,
which were 234 Myr and 262 Myr, respectively (G+14; \citealt{corr+14}).   
We suggest that the earlier age spread estimates from the overall FWHM
values should be regarded as more coarse estimates than the current
ones, since the latter account for the effects of wide ranges of
stellar rotation rates and also allow for a relative scaling of the
numbers of stars at different ages.

Finally, we emphasize that our interpretations mentioned above refer
  specifically to a comparison of the observed data with predictions of the
  current {\sc syclist} models. Given the complex evolutionary and geometric
  effects of rotation on the spectral energy distribution of stars, it will be
  important to compare our results with other, future models that incorporate
  stellar rotation.  

\section{Concluding Remarks}
\label{s:conc}

In the context of the question of the nature of eMSTOs in young and
intermediate-age star clusters in the Magellanic Clouds, we use Monte Carlo
simulations with the {\sc syclist} isochrone models to conduct a detailed
investigation of the MSTO morphologies produced by ranges in stellar rotation
rates. In doing so, we confirm a recent finding of \citet{nied+15b} that a
distribution of stellar rotation velocities yields an extent of the MSTO that
is proportional to the age of the cluster up to an age of $\sim$\,1.0 Gyr, as
observed.  
However, we find that wide ranges of stellar rotation rates yield pseudo-age
distributions (derived from cross-cuts across MSTOs in a direction
perpendicular to non-rotating isochrones for ages around those of the
clusters) that are generally narrower than those observed for clusters
at a given age. 

We compare the simulations with high-quality CMDs of NGC\,1987 and NGC\,2249,
two massive star clusters in the LMC with an age of $\sim$\,1 Gyr, close to the
oldest age for which the MSTO still consists of stars with $\cM \geq
1.7\;M_{\odot}$, which is the lowest stellar mass modeled in the
current 
{\sc syclist} models. We find that the distribution of stars across the eMSTOs
of these clusters cannot be explained solely by a distribution of stellar
rotation velocities, unless the orientations of rapidly rotating stars are
heavily biased towards an equator-on configuration. In contrast, a combination 
of distributions of stellar rotation velocities \emph{and} of stellar ages
naturally provides a good fit to the observed eMSTO morphologies. The
presence of a wide distribution of stellar rotation velocities diminishes the
range in age required to fit the eMSTO morphology of these clusters by
$\la 20$\% relative to results of previous studies that used
non-rotating isochrones. 

Even though several recent studies have provided important evidence in
support of the stellar rotation scenario for the nature of eMSTOs in
intermediate-age clusters in the Magellanic Clouds, our results suggest that
age spreads of order a few $10^8$ yr still seem to be required to fit the
detailed morphologies of the eMSTOs in those clusters, especially for the
more massive clusters among them (see Sect.~\ref{sub:rotscenario}.2). 
This may seem unexpected, given that a recent study of the very massive
cluster W3 ($\cM \sim 10^8\;M_{\odot}$, age $\sim$\,600 Myr, effective radius
$r_{\rm eff}$ = 17.5 pc) in the merger 
remnant galaxy NGC\,7252 did not find any evidence for an extended star
formation history \citep{cabr+16}. We suggest that this paradox might 
be explained if the ability of clusters to accrete and/or retain gas within
its potential well during the first few $10^8$ yr after their creation has a
significant dependence on its environment, as shown by \citet{conspe11}. Gas
present in a cluster formed in a violent dissipative merger of
$\sim$\,equal-mass galaxies like NGC 7252 \citep[e.g.,][]{whit+93},
involving collisions with relative velocities of hundreds of \kms, is likely
to be stripped rather efficiently \citep[see][]{conspe11}. In contrast,
stripping was likely much less efficient during the relatively quiescent past
evolution of dwarf galaxies such as the Magellanic Clouds, thus perhaps
allowing the accumulation of accreted or retained gas during the early
evolution of massive clusters. 

\acknowledgments
We thank the anonymous referee for thoughtful and useful comments on the manuscript.  
We are grateful to Sylvia Ekstr\"om for promptly answering our questions about 
the {\sc syclist} models. 
Support for this project was provided in part by
NASA through grant HST-GO-14174 from the Space Telescope Science Institute, 
which is operated by the Association of   Universities for
  Research in Astronomy, Inc., under NASA contract NAS5-26555. 
We acknowledge the use of the R Language for Statistical Computing, see
\url{http://www.R-project.org}.

%

\vspace{5mm}
\facility{\emph{Facilities:} HST(ACS and WFC3)}


\begin{thebibliography}{}
%
\bibitem[Bastian \& de Mink(2009)]{basdem09}
Bastian, N., \& de Mink, S. E.\ 2009, \mnras, 398, L11
\mybibitem{Bastian et al.}{2013}{bast+13}
Bastian, N., Cabrera-Ziri, I., Davies, B., \& Larsen, S. S.\ 2013, \mnras, 436,
 2852 
\mybibitem{Bastian et al.}{2016}{bast+16}
Bastian, N., Niederhofer, F., Kozhurina-Platais, V., et al.\ 2016, \mnras, 460,
 L20 
\mybibitem{Bastian et al.}{2017}{bast+17}
Bastian, N., Cabrera-Ziri, I., Niederhofer, F., et al.\ 2017, \mnras,
 465, 4795
\mybibitem{Bekki \& Mackey}{2009}{bekmac09}
Bekki, K., \& Mackey, A. D.\ 2009, \mnras, 394, 124
\mybibitem{Brandt \& Huang}{2015}{brahua15}
Brandt, T. D., \& Huang, C. X.\ 2015, \apj, 807, 25
\mybibitem{Cabrera-Ziri et al.}{2014}{cabr+14}
Cabrera-Ziri, I., Bastian, N., Davies, B., et al.\ 
 2014, \mnras, 441, 2754
\mybibitem{Cabrera-Ziri et al.}{2016}{cabr+16}
Cabrera-Ziri, I., Bastian, N., Hilker, M., et al.\ 
 2016, \mnras, 457, 809
\bibitem[Carretta et al.(2010)]{carr+10}
Carretta, E., Bragaglia, A., Gratton, R. G.\, et al.\ 2010, \aap, 505, 117
\mybibitem{Castelli \& Kurucz}{2003}{caskur03}
Castelli, F., \& Kurucz, R. L.\ 2003, in ``Modeling of Stellar Atmospheres'',
 IAU Symp.\ No.\ 210, eds. N. Piskunov et al., Poster A20
 (arXiv:astro-ph/0405087) 
\mybibitem{Claret}{2000}{clar00}
Claret, A.\ 2000, \aap, 363, 1081
\bibitem[Conroy \& Spergel(2011)]{conspe11}
Conroy, C., \& Spergel, D. N.\ 2011, \apj, 726, 36
\mybibitem{Correnti et al.}{2014}{corr+14}
Correnti, M., Goudfrooij, P., Kalirai, J. S., et al.\ 
 2014, \apj, 793, 121
\mybibitem{Correnti et al.}{2015}{corr+15}
Correnti, M., Goudfrooij, P., Puzia, T. H., \& de Mink, S. E.\ 
 2015, \mnras, 450, 3054
\mybibitem{Correnti et al.}{2017}{corr+17}
Correnti, M., Goudfrooij, P., Bellini, A., Kalirai, J. S., \& Puzia, T. H.\ 
 2017, \mnras, 467, 3628
\mybibitem{Corsaro et al.}{2017}{cors+17}
Corsaro, E., Lee, Y.-N., Garcia, R. A., et al.\ 2017, NatAs, 1, 64
\mybibitem{D'Antona et al.}{2015}{dant+15}
D'Antona, F., Di Criscienzo, M., Decressin, T., Milone, A. P., Vesperini, E., \&
 Ventura, P.\ 2015, \mnras, 453, 2637 
\bibitem[Decressin et al.(2007)]{decr+07}
Decressin, T., Meynet, G., Charbonnel, C., Prantzos, N., \& Ekstr\"om, S.\
 2007, \aap, 464, 1029
\bibitem[de Mink et al.(2009)]{demi+09}
de Mink, S. E., Pols, O. R., Langer, N., \& Izzard, R. G.\ 2009, \aap,
 507, L1
\mybibitem{Ekstr\"om et al.}{2012}{ekst+12}
Ekstr\"om, S., Georgy, C., Eggenberger, P., et al.\ 2012, \aap, 537, A146
\mybibitem{Espinosa Lara \& Rieutord}{2011}{esprie11}
Espinosa Lara, F., \& Rieutord, M.\ 2011, \aap, 533, 43
\mybibitem{For \& Bekki}{2017}{forbek17}
For, B.-Q., \& Bekki, K.\ 2017, \mnras, in press (arXiv:1703.0266) 
\mybibitem{Georgy et al.}{2013}{geor+13}
Georgy, C., Ekstr{\"o}m, S., Granada, A., et al.\ 2013, \aap, 553, A24
\mybibitem{Georgy et al.}{2014}{geor+14}
Georgy, C., Granada, A., Ekstr{\"o}m, S., et al.\ 2014, \aap, 566, A21
\bibitem[Girardi et al.(2009)]{gira+09}
Girardi, L., Rubele, S., \& Kerber, L.\ 2009, \mnras, 394, L74
\bibitem[Girardi et al.(2011)]{gira+11}
Girardi, L., Eggenberger, P., \& Miglio, A.\ 2011, \mnras, 412, L103
\mybibitem{Girardi et al.}{2013}{gira+13}
Girardi, L., Goudfrooij, P., Kalirai, J. S., et al.\ 2013, \mnras, 431, 3501
\bibitem[Glatt et al.(2008)]{glat+08}
Glatt, K., Grebel, E. K., Sabbi, E., et al.\ 2008, \aj, 135, 1703
\bibitem[Goudfrooij et al.(2009)]{goud+09}
Goudfrooij, P., Puzia, T. H., Kozhurina-Platais, V., \& Chandar, R.\ 2009,
 \aj, 137, 4988
\bibitem[Goudfrooij et al.(2011a)]{goud+11a}
Goudfrooij, P., Puzia, T. H., Chandar, R., \& Kozhurina-Platais, V.\ 2011a,
 \apj, 737, 4
\bibitem[Goudfrooij et al.(2011b)]{goud+11b}
Goudfrooij, P., Puzia, T. H., Kozhurina-Platais, V., \& Chandar, R.\ 2011b,
 \apj, 737, 3
\mybibitem{Goudfrooij et al.}{2014}{goud+14}
Goudfrooij, P., Girardi, L., Kozhurina-Platais, V., et al.\ 2014, 
 \apj, 797, 35 (G+14)
\mybibitem{Goudfrooij et al.}{2015}{goud+15}
Goudfrooij, P., Girardi, L., Rosenfield, P., et al.\ 2015, 
 \mnras, 450, 1693
\mybibthree{Gratton, Carretta, \& Bragaglia}{Gratton et al.}{2012}{grat+12}
Gratton, R., Carretta, E., \& Bragaglia, A.\ 2012, \aapr, 20, 50
\mybibitem{Huang \& Gies}{2006}{huagie06}
Huang, W., \& Gies, D. R.\ 2006, \apj, 648, 580 (HG06)
\mybibthree{Huang, Gies, \& McSwain}{Huang et al.}{2010}{huan+10}
Huang, W., Gies, D. R., \& McSwain, M. V.\ 2010, \apj, 722, 605 (HGM10)
\bibitem[Keller et al.(2011)Keller, Mackey, \& Da Costa]{kell+11}
Keller, S. C., Mackey, A. D., \& Da Costa, G. S.\ 2011, \apj, 731, 22
\mybibitem{Laidler et al.}{2008}{laid+08}
Laidler, V.\ et al.\ 2008, ``Synphot Data User's Guide'' (Baltimore: STScI)
\mybibitem{Mackey \& Broby Nielsen}{2007}{macbro07}
Mackey, A. D., \& Broby Nielsen, P.\ 2007, \mnras, 379, 151
\bibitem[Mackey et al.(2008)]{mack+08a}
Mackey, A. D., Broby Nielsen, P., Ferguson, A. M. N., \& Richardson,
 J. C.\ 2008, \apj, 681, L17
\bibitem[Milone et al.(2009)]{milo+09}
Milone, A. P., Bedin, L. R., Piotto, G., \& Anderson, J.\ 2009, \aap, 497, 755
\mybibitem{Milone et al.}{2013}{milo+13}
Milone, A. P., Bedin, L. R., Cassisi, S., et al.\ 2013, \aap, 555, A143
\mybibitem{Milone et al.}{2015}{milo+15}
Milone, A. P., Bedin, L. R., Piotto, G., et al.\ 
 2015, \mnras, 450, 3750
\mybibitem{Milone et al.}{2016}{milo+16}
Milone, A. P., Marino, A. F., D'Antona, F., et al.\ 
 2016, \mnras, 458, 4368
\mybibitem{Milone et al.}{2017}{milo+17}
Milone, A. P., Marino, A. F., D'Antona, F., et al.\ 2017, \mnras, 465, 4363
\mybibitem{Mowlavi et al.}{2012}{mowl+12}
Mowlavi, N., Eggenberger, P., Meynet, G., et al.\ 2012, \aap, 541, A41
\mybibitem{Niederhofer et al.}{2015a}{nied+15a}
Niederhofer, F., Hilker, M., Bastian, N., \& Silva-Villa, E.\ 2015a, \aap, 575, 62
\mybibitem{Niederhofer et al.}{2015b}{nied+15b}
Niederhofer, F., Georgy, C., Bastian, N.., \& Ekstr\"om, S.\ 2015b, \mnras, 453, 2070
\mybibitem{Piatti \& Bastian}{2016}{piabas16}
Piatti, A. E., \& Bastian, N.\ 2016, \mnras, 463, 1632
\bibitem[Rubele et al.(2011)]{rube+11}
Rubele, S., Girardi, L., Kozhurina-Platais, V., Goudfrooij, P., \& Kerber, L.\
 2011, \mnras, 414, 2204
\bibitem[Rubele et al.(2013)]{rube+13}
Rubele, S., Girardi, L., Kozhurina-Platais, V., et al.\ 
 2013, \mnras, 433, 2774
\bibitem[Rubele et al.(2010)]{rube+10}
Rubele, S., Kerber, L., \& Girardi, L.\ 2010, \mnras, 403, 1156
\mybibitem{Ventura et al.}{2001}{vent+01}
Ventura, P., D'Antona, F., Mazzitelli, I., \& Gratton, R.\ 2001, \apj, 550, L65
\mybibitem{Whitmore et al.}{1993}{whit+93}
Whitmore, B. C., Schweizer, F., Leitherer, C., Borne, K., \& Robert, C.\
 1993, \aj, 106, 1354
\end{thebibliography}
\end{document}